\documentclass[12pt]{spieman}  
\usepackage{amsmath,amsfonts,amssymb}
\usepackage{graphicx}
\usepackage{xcolor}
\usepackage{setspace}
\usepackage{tocloft}
\usepackage{bm}
\usepackage{ulem}
\renewcommand{\emph}[1]{\textit{#1}}

\newcommand{\eg}{\textit{e.g.}}
\newcommand{\ie}{\textit{i.e.}}

\newcommand{\etc}{\textit{etc.}}
\newcommand{\etal}{\textit{et al.}}

\newcommand{\spitz}{\textit{Spitzer}}

\newcommand{\hersc}{\textit{Herschel}}
\newcommand{\planck}{\textit{Planck}}

\newcommand{\sbs}[1]{SBS\;#1}

\newcommand{\izw}{I\;Zw\;18}

\newcommand{\mic}{\mu\textnormal{m}}

\newcommand{\emic}{\;\mic}

\newcommand{\tmic}{$\mic$}

\newcommand{\arcdeg}[2]{\ifthenelse{\equal{#2}{}}%
                         {#1$^{\circ}$}%
                         {#1$.\!\!^{\circ}$#2}}
\newcommand{\arcmin}[2]{\ifthenelse{\equal{#2}{}}%
                         {#1$^{\prime}$}%
                         {#1$.\!\!^{\prime}$#2}}
\newcommand{\arcsec}[2]{\ifthenelse{\equal{#2}{}}%
                         {#1$^{\prime\prime}$}%
                         {#1$.\!\!^{\prime\prime}$#2}}

\newcommand{\cii}{C\,\textsc{ii}}

\newcommand{\hi}{H\,\textsc{i}}

\newcommand{\oi}{O\,\textsc{i}}

\newcommand{\siii}{S\,\textsc{iii}}

\newcommand{\siII}{Si\,\textsc{ii}}

\newcommand{\hiline}{[\hi]$_{21\,\textnormal{cm}}$}

\newcommand{\ciiline}{[\cii]$_{158\mu\textnormal{m}}$}

\newcommand{\niiline}{[N\,\textsc{ii}]$_{122\mu\textnormal{m}}$}

\newcommand{\oiline}{[\oi]$_{63\mu\textnormal{m}}$}
\newcommand{\oilineb}{[\oi]$_{145\mu\textnormal{m}}$}
\newcommand{\oiiiline}{[O\,\textsc{iii}]$_{88\mu\textnormal{m}}$}

\newcommand{\siiilineb}{[\siii]$_{33.48\mu\textnormal{m}}$}

\newcommand{\siIIline}{[\siII]$_{34.82\mu\textnormal{m}}$}

\newcommand{\sms}[1]{{\mbox{{\scriptsize #1}}}}

\newcommand{\E}[1]{\times10^{#1}}

\providecommand{\refeq}[1]{\hyperref[#1]{Eq.\ (\ref{#1})}}
\newcommand{\refeqp}[1]{\hyperref[#1]{(Eq.\ \ref{#1})}}
\newcommand{\refeqnp}[1]{\hyperref[#1]{Eq.\ \ref{#1}}}
\newcommand{\refeqs}[2]{\ifthenelse{\equal{#2}{}}%
                      {\hyperref[#1]{Eqs.\ (\ref{#1})}}%
                      {\hyperref[#1]{Eqs.\ (\ref{#1})}~--~\hyperref[#2]{(\ref{#2})}}}
\newcommand{\refeqsnp}[2]{\ifthenelse{\equal{#2}{}}%
                        {\hyperref[#1]{Eqs.\ \ref{#1}}}%
                        {\hyperref[#1]{Eqs.\ \ref{#1}}~--~\hyperref[#2]{\ref{#2}}}}

\newcommand{\reftab}[1]{\hyperref[#1]{Table \ref{#1}}}
\newcommand{\reftabs}[2]{\ifthenelse{\equal{#2}{}}%
                       {\hyperref[#1]{Tables \ref{#1}}}%
                       {\hyperref[#1]{Tables \ref{#1}}~--~\hyperref[#2]{\ref{#2}}}}

\newcommand{\reffig}[1]{\hyperref[#1]{Fig.\ \ref{#1}}}
\newcommand{\reffigs}[2]{\ifthenelse{\equal{#2}{}}%
                       {\hyperref[#1]{Figs.\ \ref{#1}}}%
                       {\hyperref[#1]{Figs.\ \ref{#1}}~--~\hyperref[#2]{\ref{#2}}}}
\newcommand{\refsubfig}[2]{\hyperref[#1]{Fig.\ \ref{#1}.#2}}

\newcommand{\refchap}[1]{\hyperref[#1]{Chap.\ \ref{#1}}}
\newcommand{\refchaps}[2]{\ifthenelse{\equal{#2}{}}%
                       {\hyperref[#1]{Chaps.\ \ref{#1}}}%
                       {\hyperref[#1]{Chaps.\ \ref{#1}}~--~\hyperref[#2]{\ref{#2}}}}

\newcommand{\refsec}[1]{\hyperref[#1]{Sect.\ \ref{#1}}}
\newcommand{\refsecs}[2]{\ifthenelse{\equal{#2}{}}%
                       {\hyperref[#1]{Sects.\ \ref{#1}}}%
                       {\hyperref[#1]{Sects.\ \ref{#1}}~-- \hyperref[#2]{\ref{#2}}}}
\newcommand{\refS}[1]{\hyperref[#1]{\S\ref{#1}}}

\newcommand{\refapp}[1]{\hyperref[#1]{Appendix \ref{#1}}}
\newcommand{\refapps}[2]{\ifthenelse{\equal{#2}{}}%
                       {\hyperref[#1]{Appendices \ref{#1}}}%
                       {\hyperref[#1]{Appendices \ref{#1}}~--~\hyperref[#2]{\ref{#2}}}}

\newcounter{textlistctr}

\DeclareGraphicsExtensions{.png,.pdf,.jpg}

\title{The PRIMA promise of deciphering interstellar dust evolution with observations of the nearby Universe}

\author[a,*]{Frédéric \textsc{Galliano}}
\author[b]{Maarten \textsc{Baes}}
\author[a]{Léo \textsc{Belloir}}
\author[c]{Simone \textsc{Bianchi}}
\author[d]{Caroline \textsc{Bot}}
\author[e]{Francesco \textsc{Calura}}
\author[e]{Viviana \textsc{Casasola}}
\author[b]{Jérémy \textsc{Chastenet}}
\author[f]{Christopher \textsc{Clark}}
\author[d]{Lucie \textsc{Correia}}
\author[b]{Ilse \textsc{de Looze}}
\author[g]{Mika \textsc{Juvela}}
\author[h]{Hidehiro \textsc{Kaneda}}
\author[i]{Stavroula \textsc{Katsioli}}
\author[j,k,l]{Francisca \textsc{Kemper}}
\author[a]{Vianney \textsc{Lebouteiller}}
\author[a]{Suzanne \textsc{Madden}}
\author[m]{Mikako \textsc{Matsuura}}
\author[n]{Takashi \textsc{Onaka}}
\author[b]{Lara \textsc{Pantoni}}
\author[e]{Francesca \textsc{Pozzi}}
\author[o]{Monica \textsc{Relaño Pastor}}
\author[a]{Marc \textsc{Sauvage}}
\author[m]{Matthew \textsc{Smith}}
\author[e]{Vidhi \textsc{Tailor}}
\author[p,q]{Tsutomu T.\ \textsc{Takeuchi}}
\author[i]{Emmanuel \textsc{Xilouris}}
\author[r]{Nathalie \textsc{Ysard}}
\affil[a]{Université Paris-Saclay, Université Paris Cité, CEA, CNRS, AIM, 91191, 
          Gif-sur-Yvette, France}
\affil[b]{Sterrenkundig Observatorium, Universiteit Gent, Krijgslaan 281 S9, B-9000 Gent, 
          Belgium}
\affil[c]{INAF-Osservatorio Astrofisico di Arcetri, L. E. Fermi 5, 50125 Firenze, Italy}
\affil[d]{Université de Strasbourg, CNRS, Observatoire astronomique de Strasbourg, UMR 
          7550, F-67000 Strasbourg, France}
\affil[e]{INAF – Istituto di Radioastronomia, Via P. Gobetti 101, 40129 Bologna, Italy}
\affil[f]{Space Telescope Science Institute, 3700 San Martin Drive, Baltimore, MD 
          21218-2463, USA}
\affil[g]{Department of Physics, PO Box 64, University of Helsinki, 00014 Helsinki, 
          Finland}
\affil[h]{Graduate School of Science, Nagoya University, Furo-cho, Chikusa-ku, Nagoya 
          464-8602, Japan}
\affil[i]{National Observatory of Athens, IAASARS, GR-15236, Athens, Greece}
\affil[j]{Institute of Space Sciences (ICE), CSIC, Can Magrans, 08193 Cerdanyola del 
         Vallés, Barcelona, Spain}
\affil[k]{ICREA, Pg. Lluís Companys 23, Barcelona, Spain}
\affil[l]{Institut d’Estudis Espacials de Catalunya (IEEC), E-08034 Barcelona, Spain}
\affil[m]{School of Physics and Astronomy, Cardiff University, The Parade, 
          Cardiff CF24 3AA, UK}
\affil[n]{Department of Astronomy, Graduate School of Science, University of Tokyo, 7-3-1 
          Hongo, Bunkyo-ku, Tokyo 113-0033, Japan}
\affil[o]{Dept. Física Teórica y del Cosmos, Universidad de Granada, Spain}
\affil[p]{Division of Particle and Astrophysical Science, Nagoya University,
          Furo-cho, Chikusa-ku, Nagoya 464-8602, Japan}
\affil[q]{The Research Center for Statistical Machine Learning, the Institute of 
          Statistical Mathematics, 10–3 Midori-cho, Tachikawa, Tokyo 190–8562, Japan}
\affil[r]{IRAP, Université de Toulouse, CNRS, UPS, IRAP, Toulouse Cedex 4, France}

\cftpagenumbersoff{figure}
\cftpagenumbersoff{table} 
\begin{document} 
\maketitle

\begin{abstract}
This paper develops a few science cases, using the PRIMA far-IR probe, aimed at achieving several breakthroughs in our understanding of the dust properties and their evolution.
We argue that the specific observational capabilities of PRIMA, namely its unprecedented sensitivity over the whole far-IR range and the possibility to obtain continuous spectra between $\lambda=24$ and 235~\tmic, are essential to progress in our understanding of the physics of the interstellar medium and galaxy evolution.
Our science cases revolve around observations of nearby galaxies.
We discuss the importance of detecting the IR emission of the diffuse interstellar medium
of these galaxies, including very low-metallicity systems.
We also discuss the opportunity of detecting various solid-state features to understand the mineralogy of interstellar grains.
Finally, we stress the unique opportunity brought by the possible simultaneous measures of both the dust continuum and the far-IR fine-structure gas lines.
These science cases could be distributed in a few large programs.
\end{abstract}


{\noindent \footnotesize\textbf{*}Frédéric GALLIANO, \linkable{frederic.galliano@cea.fr} }

\begin{spacing}{1}


\section{Introduction}
\label{sec:intro}

Dust grains are a crucial ingredient of the physics of the \textit{InterStellar Medium} (ISM), obscuring starlight, heating the neutral gas, and catalyzing numerous prominent chemical reactions \cite{galliano18}.
Dust observables also provide invaluable diagnostics of the physical conditions from Galactic clouds to high redshift galaxies, allowing the observer to derive \textit{Star Formation Rates} (SFR), dust and gas masses, the magnetic field orientation, \etc\
The knowledge of interstellar dust physics is thus necessary for understanding both galaxy evolution \cite{dubois24} and the process of star formation, down to the formation of planetary systems \cite{lebreuilly24} and prebiotic chemistry \cite{guelin22}.
This knowledge is unfortunately hampered by large uncertainties.
First, the precise dust constitution (chemical composition, structure, size and shape distributions) in a given region is extremely difficult to infer, due to degeneracies between the optical properties and the size distribution.
In addition, the grain properties are known to evolve with the density and UV field, in the \textit{Milky Way} (MW) \cite{ysard15}.
Dust observables from the UV to the microwave regime exhibit complex dependencies when spatially resolved at scales of interstellar clouds \cite{lequeux82,valencic03,gordon03,maiz-apellaniz12,gordon24}.
Outside the MW, these parameters are also important, but the abundance of elements heavier than Helium, the \textit{metallicity}, which is quite uniform in the MW, appears to be the main quantity shaping the dust observables of galaxies of different types \cite{sandstrom12,galliano21,whitcomb24}.

Our current poor knowledge of the dust properties and their evolution is thus an obstacle to answering fundamental questions in astrophysics.
For instance, the dust-to-gas mass ratio of galaxies as a function of their metallicity, a relation essential to quantifying the balance between dust formation and destruction, differs significantly in the low-metallicity regime if it is measured using the thermal emission observed in the  \textit{Far-InfraRed} (FIR) or using elemental depletions from UV absorption lines \cite{roman-duval21,roman-duval22a,roman-duval22b}.
This discrepancy might be linked to both our lack of knowledge of the FIR opacity of grain mixtures at low metallicity, and to biases of the depletion method.
Another example of the limitations we face is the so-called \textit{submillimeter (submm) excess}, a systematic deviation between the submm-mm observations and contemporary state-of-the art models \cite{galliano03,galliano05,galametz09,bot10,paradis11,planck-collaboration11,dale12}.
This excess is more prominent at low metallicity and potentially biases dust mass estimates.
Finally, the UV-to-FIR extinction law also depends on the environment. 
Its environmental variation is relatively well-known in the MW, from an empirical point of view \cite{fitzpatrick19,gordon23}, although its interpretation in terms of variation of the grain constitution is still debated.
However, outside the MW, our knowledge of the extinction curve is limited to a handful of sightlines in the most nearby galaxies, primarily the Magellanic Clouds \cite{gordon03,gordon24}.
The Magellanic extinction curves do not resemble those of the MW.
Magellanic extinction curves are sometimes used to unredden observations of galaxies without any rigorous justification.
This potentially induces systematic effects in all the UV-to-mid-IR (UV-to-MIR) tracers.

\textit{Ab initio} simulations of the formation and evolution of interstellar grains are not yet feasible \cite{draine09}.
We thus essentially rely on an empirical approach, complemented by our knowledge of solid-state physics.
There are several observational clues that we would need to obtain to move forward.
First, contemporary dust models are solely constrained by the IR \textit{Spectral Energy Distribution} (SED), the UV-to-MIR extinction curve, the elemental depletions and the polarization fraction in extinction and emission of the diffuse ISM of the MW \cite{hensley23,ysard24}.
This heterogeneous set of constraints is required to solve the degeneracies between the optical properties and the size distribution \cite{galliano18}.
We thus lack an extragalactic dust model that could account for the effects of metallicity, as well as those of the density and UV field.
Ideally, we would need similar sets of constraints in the diffuse ISM of external galaxies, starting with the Magellanic clouds, to thoroughly understand the effect of metallicity.
In addition, we would need a way to quantify the evolution of the diffuse ISM dust mixture into denser and more illuminated regions, where these observables are more difficult to obtain and more ambiguous to interpret due to the mixing of physical conditions along the sightline.
Second, the low-metallicity regime is crucial to understanding the early stages of dust formation and inferring what happens in primordial galaxies \cite{palla24}.
Low-metallicity systems have been studied with \hersc\ \cite{madden13,sandstrom13,remy-ruyer14,chastenet17,galliano21}, but we lack statistics at \textit{extremely} low metallicity ($Z\lesssim1/10\;Z_\odot$) and at quiescent star-formation activities.
Finally, MIR solid-state features in emission or absorption are instrumental in characterizing accurate stoichiometry and structure of the material constituting the grains.
JWST, after ISO and \spitz, is currently detecting numerous such features, but it is spectrally limited to the range where they are mixed with the bright features coming from \textit{Polycyclic Aromatic Hydrocarbons} (PAHs).
Yet, there are numerous features longward of $\lambda=28$~\tmic\ that would allow us to better characterize the grain composition.
PRIMA offers the promise to address several of these issues, thanks to its unmatched sensitivity in the FIR, its continuous-wavelength spectroscopic capability, and its polychromatic polarization imager \cite{glenn21,moullet23}.

The present paper proposes ways to address the limitations listed above using PRIMA observations of nearby galaxies.
We focus our discussion on what PRIMA data will bring to our science questions.
However, most of our targets have been fully mapped with \spitz\ and some are currently being imaged with the JWST.
These mid-IR data will be instrumental in complementing PRIMA's spectral coverage at short wavelengths.
In \refsec{sec:diffuse}, we argue for the necessity to obtain deep observations of the diffuse ISM of the Magellanic Clouds.
We then discuss the constraints on interstellar grain mineralogy accessible with PRIMA, in \refsec{sec:mineralogy}.
The degeneracy between the effects of metallicity and star-formation activity, and a way to solve it, is presented in \refsec{sec:LSB}.
Finally, \refsec{sec:gasndust} proposes a way of combining dust and gas tracers to better understand the grain properties of dense, UV-illuminated regions.
We end with a conclusion in \refsec{sec:concl}.
Throughout this paper, we use \textit{The Heterogenous dust Evolution Model for Interstellar Solids} (\texttt{THEMIS}) \cite{jones17} as a reference.
All the figures and flux estimates are done using this model.
Our observing strategy is based on the formulae summarized in \refapp{app:sens}.


\section{Unveiling the Elusive Dust Properties of the Diffuse ISM of Nearby 
           Galaxies}
\label{sec:diffuse}

We have stated in \refsec{sec:intro} that the constitution of interstellar dust in external galaxies is believed to differ significantly from the MW. 
Dust build-up and evolution indeed depend on the particular \textit{Star Formation History} (SFH) of each galaxy.
In particular, the metallicity, $Z$, appears to be one of the most important factors \cite{galliano21}. 
This parameter, $Z$, quantifies the cumulated elemental enrichment of a system. 
To properly interpret observations of galaxies, we thus need to understand how the dust properties vary as a function of $Z$. 
Yet, as we have seen, contemporary dust models \cite{jones17,hensley23}, that are used to provide such an interpretation, are exclusively constrained by observations of the MW, a system with a narrow $Z$ range around the Solar value, $Z_\odot$. 
We are therefore biased by the particular properties of the MW when modeling the dust SED of other galaxies. 
This bias especially questions our ability to accurately understand nearby dwarf galaxies and early Universe galaxies for which the metal enrichment is expected to be low.

  \subsection{The need for a fully-constrained extragalactic dust model}

Arguably, no properly-constrained dust model of external galaxies currently exists. 
This is because there is a deficit of observational constraints. 
Several teams have fitted the extinction curves of the Magellanic Clouds with realistic optical properties and a range of grain sizes \cite{pei92,weingartner01,clayton03}.
There are however strong degeneracies between these optical properties and the size distribution.
Contemporary dust models for the Milky Way address this degeneracy by combining the constraints coming from the IR emission, the UV-NIR extinction, and elemental depletions \cite{zubko04,draine07,jones17,siebenmorgen20} and more recently the polarization \cite{hensley23,ysard24}.
The mixing of physical conditions along the sightline and within the telescope beam indeed renders the SED degenerate. 
For instance, if we are observing a region where there is a gradient of \textit{InterStellar Radiation Field} (ISRF), we will not be able to distinguish an overabundance of small grains from the spread due to the variation of the equilibrium temperatures of large grains (\eg\ Fig.~3 of Galliano \etal, 2018\cite{galliano18}). 
This is why dust models are calibrated on observations of the diffuse ISM of the MW. 
The low optical depth of this medium ($A(\textnormal{V})\simeq0.1$ for $N(\textnormal{\hi})\simeq2\E{20}$~H/cm$^2$, at $Z=Z_\odot$) ensures that the grains will be uniformly illuminated by the average ISRF. 
It is possible to build the observed SED of the diffuse ISM of the MW, by averaging the high Galactic latitude fluxes given by IRAS, COBE and \planck. 
This is however not yet possible in external galaxies. 
IRAS, COBE and \planck\ did not sufficiently resolve galaxies to allow the extraction of their diffuse ISM emission, without contaminations by denser clouds. 
And other observatories with a finer angular resolution, such as \spitz\ and \hersc, were not sensitive enough (\hersc\ could barely go below $N(\textnormal{\hi})\simeq10^{22}$~H/cm$^2$).

\begin{figure}[htbp]
  \includegraphics[width=\textwidth]{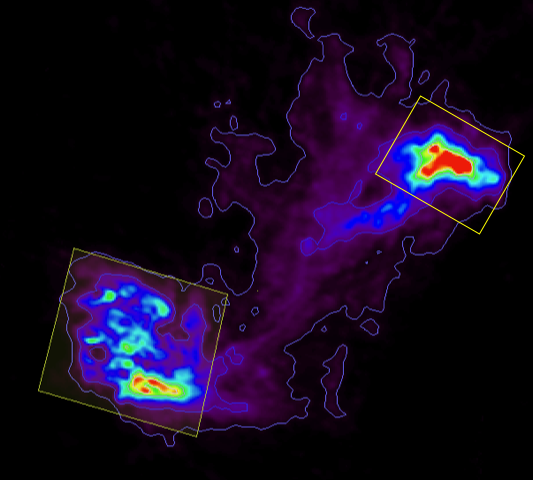}
  \caption{\hi\ map of the LMC / SMC \cite{bruens05}. 
           The contours correspond to $N(\textnormal{\hi})=10^{20}$~H/cm$^2$ and
           $N(\textnormal{\hi})=2\E{20}$~H/cm$^2$. 
           The yellow rectangles are areas covering most of the \hi\ emission of both 
           sources (LMC: $12^\circ\times12^\circ$; SMC: $8^\circ\times6^\circ$).}
  \label{fig:lmcsmc}
\end{figure}
The good angular resolution and the exceptional sensitivity of PRIMA over the whole \textit{Mid-InfraRed-to-Far-InfraRed} (MIR-to-FIR) window gives us a way to palliate this problem, without having to resort to combining observations that trace different scales (also called \textit{feathering}) \cite{clark23}. 
The two closest galaxies, the \textit{Large and Small Magellanic Clouds} (LMC \&\ SMC; distances: $d_\sms{LMC}=50$~kpc and $d_\sms{SMC}=60$~kpc; metallicities: $Z_\sms{LMC}=1/2\;Z_\odot$
and $Z_\sms{SMC}=1/5\;Z_\odot$; \reffig{fig:lmcsmc}) are ideal targets. 
We will be able to resolve regions of $\simeq10$~pc size at $\lambda=250$~\tmic. 
With the addition of already estimated elemental depletions and extinction curves \cite{gordon03,tchernyshyov15}, the well-characterized broad-band SED of their diffuse emission per H atom will allow us to build the first properly-constrained extragalactic, non-Solar metallicity, dust models.

  \subsection{Possible observing strategy}

To measure the MIR-to-FIR SED of the diffuse ISM of the LMC and SMC, we need to make deep maps of these two galaxies in all available bands. 
In addition, FIR measures of the polarization fraction are important to constrain the
grain composition \cite{guillet18}.
We could use PRIMAger, to measure both the total power and the polarization fraction
(\reftab{tab:diff}).
\begin{enumerate}
  \item We could make large maps of the LMC ($12^\circ\times12^\circ$) and SMC 
    ($8^\circ\times6^\circ$) with PRIMAger at the 4 long-wavelength bands.
  \item To keep the observing time under a reasonable duration, we could also make smaller 
    maps ($0.2^\circ\times0.2^\circ$) of a diffuse region in each galaxy, in each of the  
    short-wavelength bands, as confusion is less problematic in the MIR.
  \item Similarly, detecting the polarization fraction would be prohibitive over the whole 
    galaxies, but could be feasible for the same $0.2^\circ\times0.2^\circ$ region with a 
    deeper scanning time.
\end{enumerate}
This observing strategy is summarized in \reftab{tab:diff}.
Taking advantage of the parallel mode, this science case could be fully executed in a total of $235+325+78+235=963$~hours.
If we do not aim at detecting the polarization intensity at 96~\tmic, this total observing time would be reduced to $235+190+78+190=693$~hours.
\begin{table}[htbp]
  \caption{Sensitivity and observing time estimates of the diffuse ISM of the 
           Magellanic Clouds (\refsec{sec:diffuse}).
           Lines highlighted in grey are observations we could drop to keep the observing
           time under control.
           Taking advantage of the parallel mode observations, each one of the four blocks 
           of this table represents one single observation, 
           where all the bands are integrated in parallel.}
  \label{tab:diff}
  \centering
  \begin{tabular}{|l|l|r|r|}
    \hline
      \textbf{PRIMAger band} & \textbf{Field} & \textbf{Sensitivity} & \textbf{Integration time} \\
    \hline
      \multicolumn{4}{|c|}{\textbf{LMC (long-wavelength total power)}} \\
    \hline
      PPI1 96~\tmic\ (total power) & $12^\circ\times12^\circ$ & 3.91 MJy/sr 
        & 235 hours \\
      PPI2 126~\tmic\ (total power) & $12^\circ\times12^\circ$ & 5.12 MJy/sr 
        & 79 hours \\
      PPI3 172~\tmic\ (total power) & $12^\circ\times12^\circ$ & 4.49 MJy/sr 
        & 55 hours \\
      PPI4 235~\tmic\ (total power) & $12^\circ\times12^\circ$ & 3.00 MJy/sr 
        & 68 hours \\
    \hline
      \multicolumn{4}{|c|}{\textbf{LMC (short-wavelength \&\ long-wavelength polarization)}} \\
    \hline
      PHI1 24--45~\tmic\ & $0.2^\circ\times0.2^\circ$ & 0.21 MJy/sr & 190 hours \\
      PHI2 45--84~\tmic\ & $0.2^\circ\times0.2^\circ$ & 1.20 MJy/sr & 1.74 hours \\
      \textcolor{gray}{PPI1 96~\tmic\ (polarization)} 
        & \textcolor{gray}{$0.2^\circ\times0.2^\circ$}
        & \textcolor{gray}{$3.91\times0.1$ MJy/sr} \
        & \textcolor{gray}{325 hours} \\
      PPI2 126~\tmic\ (polarization) & $0.2^\circ\times0.2^\circ$ 
        & $5.12\times0.1$ MJy/sr & 110 hours \\
      PPI3 172~\tmic\ (polarization) & $0.2^\circ\times0.2^\circ$ 
        & $4.49\times0.1$ MJy/sr & 77 hours \\
      PPI4 235~\tmic\ (polarization) & $0.2^\circ\times0.2^\circ$ 
        & $3.00\times0.1$ MJy/sr & 92 hours \\
    \hline
      \multicolumn{4}{|c|}{\textbf{SMC (long-wavelength total power)}} \\
    \hline
      PPI1 96~\tmic\ (total power) & $8^\circ\times6^\circ$ & 3.91 MJy/sr 
        & 78 hours \\
      PPI2 126~\tmic\ (total power) & $8^\circ\times6^\circ$ & 5.12 MJy/sr 
        & 26 hours \\ 
      PPI3 172~\tmic\ (total power) & $8^\circ\times6^\circ$ & 4.49 MJy/sr 
        & 19 hours \\
      PPI4 235~\tmic\ (total power) & $8^\circ\times6^\circ$ & 3.00 MJy/sr 
        & 23 hours \\
    \hline
      \multicolumn{4}{|c|}{\textbf{SMC (short-wavelength \&\ long-wavelength polarization)}} \\
    \hline
      PHI1 24--45~\tmic\ & $0.2^\circ\times0.2^\circ$ & 0.21 MJy/sr & 190 hours \\
      PHI2 45--84~\tmic\ & $0.2^\circ\times0.2^\circ$ & 1.20 MJy/sr & 1.74 hours \\
      \textcolor{gray}{PPI1 96~\tmic\ (polarization)} 
        & \textcolor{gray}{$0.2^\circ\times0.2^\circ$}
        & \textcolor{gray}{$3.91\times0.1$ MJy/sr} \
        & \textcolor{gray}{325 hours} \\
      PPI2 126~\tmic\ (polarization) & $0.2^\circ\times0.2^\circ$ 
        & $5.12\times0.1$ MJy/sr & 110 hours \\
      PPI3 172~\tmic\ (polarization) & $0.2^\circ\times0.2^\circ$ 
        & $4.49\times0.1$ MJy/sr & 77 hours \\
      PPI4 235~\tmic\ (polarization) & $0.2^\circ\times0.2^\circ$ 
        & $3.00\times0.1$ MJy/sr & 92 hours \\
    \hline
  \end{tabular}
\end{table}

The \spitz\ and \hersc\ maps of the Magellanic clouds are good \cite{meixner06,gordon11,meixner13}. 
However, they are not deep enough for this science case. 
Besides, PRIMA will provide a finer spectral sampling.
The main challenge of this science case is the confusion with the diffuse dust emission of the MW and with the cosmic infrared background. 
The redundancy provided by a large number of pixels towards many sightlines is the key to subtract these contaminations.
This is why we need to map large areas to allow us using a statistical decomposition method \cite{auclair24}.

\subsection{Assumptions}

The flux sensitivities in \reftab{tab:diff} have been estimated using the \texttt{THEMIS} dust model \cite{jones17} (\refapp{app:sens}), with the following assumptions.
\begin{itemize}
  \item The ISRF intensity appears to be higher in the Magellanic clouds than in the Milky 
    Way, because of the presence of a higher fraction of young stars and a 
    lower ISM 
    opacity due to the lower dust-to-gas mass ratio \cite{galliano11}. 
    Galliano \etal\ (2011) \cite{galliano11} showed that $U\simeq3.5$ in the LMC, and 
    $U\simeq8$ in the most diffuse sightlines.
    To remain conservative, we assume $U=3$.
  \item We aim to reach the equivalent of $N(H)=2\E{20}$~H/cm$^2$ in the MW.
    At first approximation, the emission and the \textit{dust} column density both scale 
    with $Z$ (\refapp{app:sens}).
    We should thus scale the emission of the model by $Z/Z_\odot$.
    However, we are interested in the emission of an optically thin medium 
    ($A(V)\simeq0.1$). 
    This $A(V)\simeq0.1$ will be reached at a $Z_\odot/Z$ times higher column density than 
    in the MW. 
    $Z$ therefore cancels out in this estimate and we simply need to use the MW model with
    $N(H)=2\E{20}$~H/cm$^2$.
\end{itemize}


\section{Surveying the Mineralogical Diversity of the ISM}
\label{sec:mineralogy}

Outside the diffuse ISM, where we are looking at denser regions of the MW and nearby galaxies, the dust model constraints discussed in \refsec{sec:diffuse} become more ambiguous to interpret, because of the mixing of physical conditions along the sightline. 
In addition, extinction and depletions become more difficult to obtain, as it becomes more difficult to detect background sources in front of dense clouds.
We thus need to find other ways to study the dust properties in denser regions.

  \subsection{The need for a deep extragalactic mineralogical survey}

In particular, the chemical composition and structure of dust grains is still widely unknown.
We have approximate indirect constraints coming from the elemental depletions, in the diffuse ISM, and from the detection of a few broad features. 
The large uncertainty of dust models however is a consequence of our ignorance of this composition and structure. 
For instance, Zubko \etal\ (2004)\cite{zubko04} assume it is made of $4.6\,\%$ Polycyclic Aromatic Hydrocarbons (PAHs), $16.4\,\%$ of graphite and $79\,\%$ of amorphous silicates; 
Compiègne \etal\ (2011)\cite{compiegne11} assume it is made of $7.7\,\%$ of PAHs, $15.8\,\%$ of amorphous carbon and $76.5\,\%$ of amorphous silicates; 
Jones \etal\ (2017)\cite{jones17} assume it is made of $31.1,\%$ of partially hydrogenated amorphous carbon and $68.9\,\%$ of amorphous silicates (forsterite and enstatite in equal proportions) with hydrogenated amorphous carbon mantles and Fe/FeS inclusions; 
Hensley \&\ Draine (2022)\cite{hensley23} assume it is made of $5\,\%$ PAHs and $95\,\%$ of a mash-up of silicate, hydrocarbons, iron and various oxides. 
This uncertainty is the main limitation of dust studies and of dust-based diagnostics of the physical conditions. 
This is also the limitation in the precision of gas-physics simulations, such as \textit{PhotoDissociation Region} (PDR) models \cite{rollig07}. 

The most straightforward way to more precisely constrain the dust composition would be to look for solid-state features. 
This is the only way to unambiguously identify a particular chemical composition. 
The MIR range is potentially the richest domain as it is where the vibrational modes of chemical bonds are located. 
For instance, we know that interstellar silicates are mostly amorphous \cite{spoon22}.
However, precisely knowing this fraction and the way it varies with the physical conditions (radiation field, gas density and metallicity) would provide invaluable constraints on dust evolution. 
Another puzzle is that there is too much depleted oxygen in the ISM, compared to what we can put in silicates. 
It is thus possible that a fraction of the dust is in the form of various oxides (\eg\
Al$_2$O$_3$, CaCO$_3$, \etc) or organic carbonates \cite{jones19}. 
The bands of these compounds may have eluded previous spectroscopic surveys. 
Finally, recent X-ray investigations of the dust chemical composition in our Galaxy suggest that Mg-rich amorphous pyroxene represents the largest fraction of dust (about $70\,\%$ on average \cite{psaradaki23}). 
This fraction may change with environments and amorphous pyroxenes can be studied through their MIR-to-FIR features as well.

  \subsection{Possible observing strategy}

\begin{figure}[htbp]
  \includegraphics[width=\textwidth]{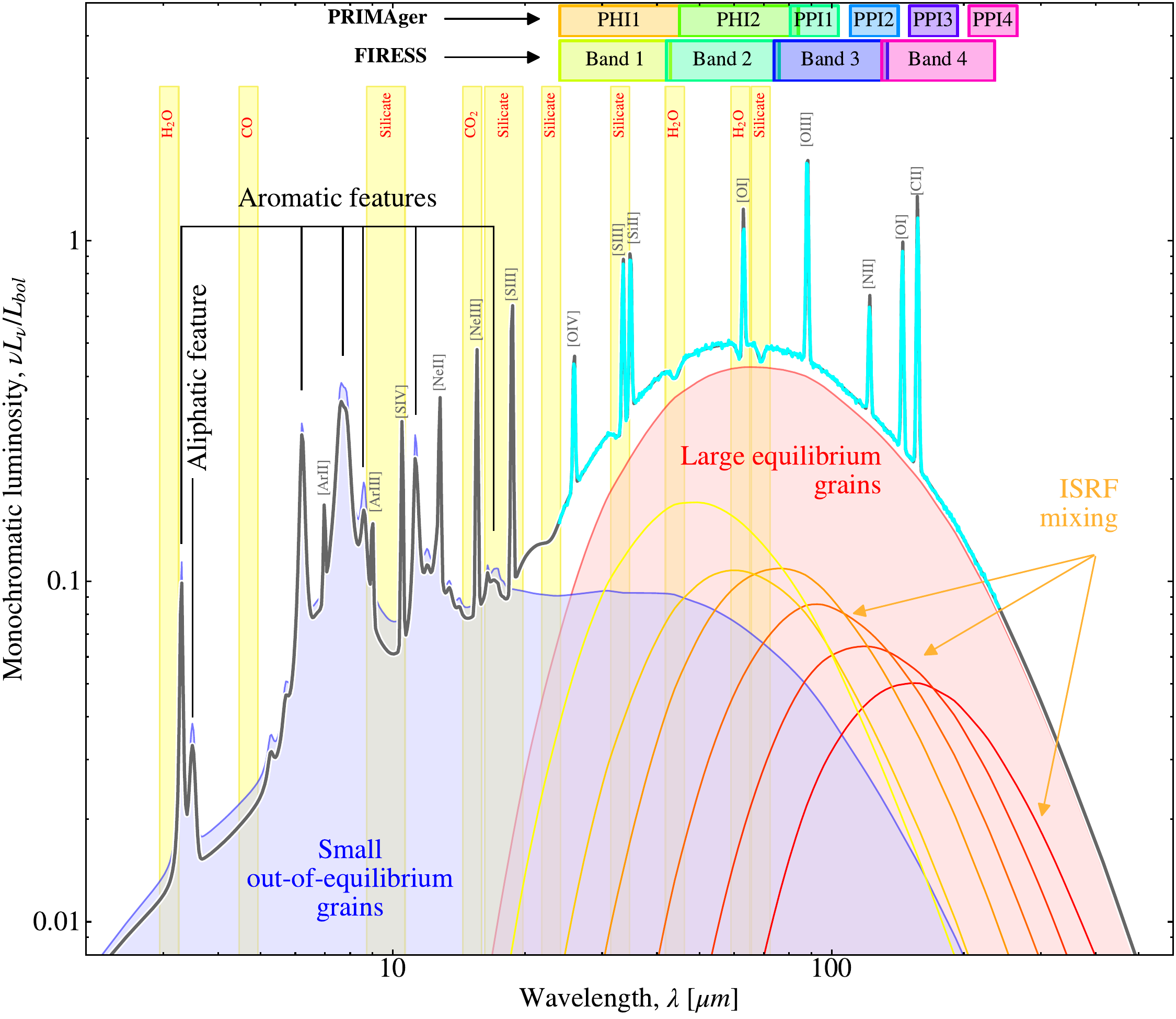}
  \caption{Typical SED of a star-forming region simulated with the THEMIS model 
           \cite{jones17}. 
           The cyan line is a simulated $R=150$ FIRESS spectrum ($100\sigma$). 
           We show a few absorption features from crystalline silicates and ices 
           \cite{van-der-tak18}, and the brightest gas lines \cite{cormier19}.}
  \label{fig:mineralogy}
\end{figure}
Low \textit{spectral resolution} ($R$), high \textit{signal-to-noise ratio} (S/N) observations over the whole MIR-to-FIR range would give us access to a wealth of features (\reffig{fig:mineralogy}). 
However, such spectra were only obtained in the past by combining ISOSWS and ISOLWS \cite{peeters02}, with a poor sensitivity, a poor spatial resolution and stitching problems. 
This is a domain where PRIMA will particularly pioneer. 
We could map star-forming regions of different metallicities, such as in the LMC and SMC. 
It would be important to obtain maps in order to understand how these features vary with the distance from the central cluster.
This would highlight the role of photo-processing as well as the role of mantle growth. 
We could be looking for both emission and absorption features. 
However, the emission features come primarily from small grains or large hot grains. 
They might not represent the bulk of the dust mass. 
This is why absorption features are preferred. 
This is also why we would need to target star-forming regions, so that we have a strong MIR
background.

We could thus use FIRESS to perform $R\simeq150$ spectral maps of a sample of 20 nearby extragalactic star-forming regions ($\simeq20^\prime\times20^\prime$), in the Magellanic clouds.
Taking advantage of the parallel observations of bands 1 and 3, and of bands 2 and 4, 20 regions could be mapped in 996~hours (\reftab{tab:mineralogy}).
\begin{table}[htbp]
  \caption{Spectroscopic time estimates for the mineralogical studies of Magellanic 
           star-forming regions.}
  \label{tab:mineralogy}
  \centering
  \begin{tabular}{|l|l|r|r|}
    \hline
      \textbf{Band [$\boldsymbol{\mu}$m]} & \textbf{Fields} 
       & \textbf{Sensitivity [W/m$^{\boldsymbol{2}}$]} & \textbf{Time [hours]} \\
    \hline
      \multicolumn{4}{|c|}{\textbf{LMC}} \\
    \hline
       Band 1 & $12\times(20^\prime\times20^\prime)$ & $8.8\E{-17}$ & 180 \\
       Band 2 & $12\times(20^\prime\times20^\prime)$ & $3.2\E{-16}$ & 13 \\
       Band 3 & $12\times(20^\prime\times20^\prime)$ & $7.0\E{-16}$ & 1.2 \\
       Band 4 & $12\times(20^\prime\times20^\prime)$ & $5.4\E{-16}$ & 0.7 \\
    \hline
      \multicolumn{4}{|c|}{\textbf{SMC}} \\
    \hline
       Band 1 & $8\times(20^\prime\times20^\prime)$ & $3.5\E{-17}$ & 748 \\
       Band 2 & $8\times(20^\prime\times20^\prime)$ & $1.3\E{-16}$ & 55 \\
       Band 3 & $8\times(20^\prime\times20^\prime)$ & $2.8\E{-16}$ & 4.9 \\
       Band 4 & $8\times(20^\prime\times20^\prime)$ & $2.2\E{-16}$ & 3.1 \\
    \hline
  \end{tabular}
\end{table}

\subsection{Assumptions}

The flux sensitivity has been estimated using the \texttt{THEMIS} dust model \cite{jones17} with the following assumptions (\refapp{app:sens}).
\begin{itemize}
  \item We assume a radiation field intensity of $U=35$ and     
    a Hydrogen column density of $N(\textnormal{\hi})=10^{22}$~H/cm$^2$. 
    This corresponds approximately to the extended parts of the star-forming regions of 
    the LMC \cite{galliano11}.
  \item We would require a high S/N ($\simeq100$), because the goal is not so much to 
    detect the intensity of a feature, but to unambiguously measure it. 
    These features are usually weak and broad, we thus need to make sure that they are not 
    diluted in the continuum.
  \item The sizes ($20^\prime\times20^\prime$) are those of typical star-forming regions 
    in the LMC / SMC (N$\,$11, N$\,$66, \etc).
\end{itemize}


\section{Disentangling the Effects of Metallicity and Star Formation Activity 
            on the Dust Properties}
\label{sec:LSB}

Our understanding of both \textit{local} (\ie\ at the scale of clouds) and \textit{global} or \textit{cosmic} (\ie\ at the scale of galaxies) dust evolution relies primarily on the empirical evidence that the grain properties are changing with the physical conditions \cite{calura08,calura17,de-looze20,galliano21}.
To that purpose, we can consider different targets as snapshots of galaxy evolution at different stages.
As mentioned in \refsec{sec:intro}, the most used environmental parameter to quantify the evolution of a galaxy is its metallicity, $Z$, as it traces the cumulated elemental enrichment of its ISM. 
However, different SFHs, and thus different evolutionary paths, can lead to the same $Z$ at a given age. 
Consequently, we often face a degeneracy between the effects of metal enrichment and star formation activity, when attempting to interpret dust evolution trends.

  \subsection{The necessity to obtain IR observations of quiescent 
                   low-metallicity galaxies}

\begin{figure}[htbp]
  \includegraphics[width=\textwidth]{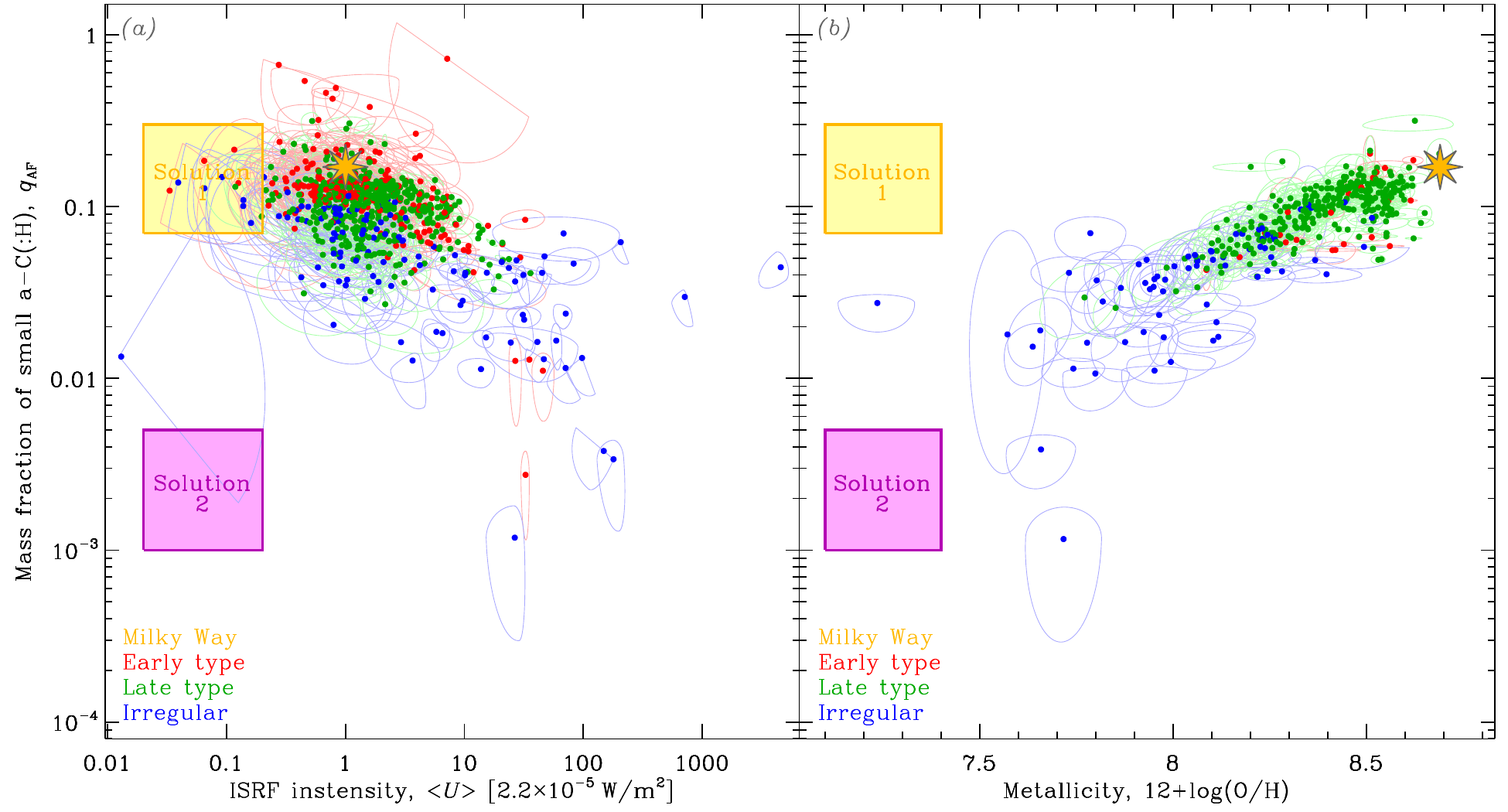}
  \caption{The potential of quiescent very-low-$Z$ galaxies. 
           We show the mass fraction of small amorphous carbon grains, $q_\sms{AF}$, 
           as a function of
           starlight intensity, $\langle U\rangle$ (panel \textit{a}), and metallicity.
           $Z$ (panel \textit{b}). 
           Each point corresponds to one galaxy of the Galliano \etal\ (2021) 
           \cite{galliano21} sample. 
           We have added two hypothetical observations of a quiescent very-low-$Z$
           galaxy (solutions 1 and 2). 
           Such observations would break the degeneracy between $Z$ and 
           $\langle U\rangle$, as they can not be consistent with both trends.}
  \label{fig:degeneracy}
\end{figure}

This degeneracy is illustrated by \reffig{fig:degeneracy}, showing the evolution of the fraction of small amorphous carbon grains as a function of $Z$ and average starlight intensity, $\langle U\rangle$, in individual galaxies \cite{galliano21}. 
It ambiguously suggests that the evolution of these small grains could be driven either by Z or by the \textit{specific Star Formation Rate} (sSFR). 
This is because the low-Z galaxies detected with Herschel are primarily actively star-forming. 
This selection effect therefore results in a correlation between these two parameters in our sample, and we are unable to understand which one is fundamental.
This degeneracy is encountered when looking at other quantities, too.

There is however a population of low sSFR, low-$Z$ galaxies, distinct from those in our
sample \cite{lara-lopez13}. 
Deriving the dust properties in these objects would thus allow us to break this type of degeneracy, as these sources would necessarily appear as a distinct branch in one of the two panels of \reffig{fig:degeneracy} (solution 1 or solution 2). In addition,
learning more about these local galaxies is interesting for the interpretation of deep
surveys, as they probe the faint end of the luminosity function. 
These objects are also local analogs to distant primordial galaxies. 
Herschel was successful in detecting a few very low-$Z$ star-forming objects (essentially \izw\ and \sbs{0335-052}). 
PRIMA could observe a sample of 100 nearby very low-$Z$ galaxies (1/10 to 1/30~$Z\odot$; typical size $2^\prime\times2^\prime$). 
For instance, the ALFALFA \hiline\ survey \cite{haynes18} contains hundreds of galaxies with $M(\textnormal{\hi})<10^8\;M_\odot$ and dozens with $M(\textnormal{\hi})<10^7\;M_\odot$. 
Among them, several have been well observed in stellar and ionized gas tracers: the 12 sources from the SHIELD sample \cite{cannon11}; the very low surface density, nearby object, Leo P \cite{giovanelli13}; the most metal-poor gas-rich galaxy known to date, AG$\,$198691 \cite{hirschauer16}. 
We also need more statistics in the low-$Z$ / high-sSFR branch. 
We thus need to explore the whole sSFR range at very low $Z$.

  \subsection{Possible observing strategy}  

The unprecedented sensitivity of PRIMA should allow us to detect these objects at all IR wavelengths. 
Obtaining the well-sampled IR SED of these objects would allow us to understand the role of Z and sSFR in shaping the dust properties at early stages.
Ideally, we would like to map this galaxy sample with all PRIMAger bands.
We would also need low-resolution spectral maps, with FIRESS, to better constrain the shape of the continuum in different regions.
Mapping 100 of these sources would take less 31 hours (\reftab{tab:LSBband}),
and $R\simeq150$ spectral mapping (except maybe in band 1) would take less than 100 hours (\reftab{tab:LSBspec}).

  \subsection{Assumptions}

We do not know much about the ISM properties of low surface brightness galaxies. 
Most of these objects have been detected only through their stellar emission or \hiline\ line.
The flux sensitivity has been estimated using the \texttt{THEMIS} dust model \cite{jones17}, assuming a typical $U=3$, similar to \refsec{sec:diffuse}, and aiming for an H column density of $N(\textnormal{\hi})=10^{21}$~H/cm$^2$.
\begin{table}[htbp]
  \caption{PRIMAger time estimates for the quiescent low-metallicity sample.}
  \label{tab:LSBband}
  \centering
  \begin{tabular}{|l|l|r|r|}
    \hline
      \textbf{Band [$\boldsymbol{\mu}$m]} & \textbf{Fields} 
        & \textbf{Sensitivity [MJy/sr]}
        & \textbf{Time [hours]} \\
    \hline
      PHI1 & $100\times(2^\prime\times2^\prime)$ & 0.85 & 31 \\
      PHI2 & $100\times(2^\prime\times2^\prime)$ & 4.9 & 0.3 \\
      PPI1 & $100\times(2^\prime\times2^\prime)$ & 16 & $<0.1$ \\
      PPI2 & $100\times(2^\prime\times2^\prime)$ & 21 & $<0.1$ \\
      PPI3 & $100\times(2^\prime\times2^\prime)$ & 18 & $<0.1$ \\
      PPI4 & $100\times(2^\prime\times2^\prime)$ & 12 & $<0.1$ \\
    \hline
  \end{tabular}
\end{table}
\begin{table}[htbp]
  \caption{FIRESS time estimates for the quiescent low-metallicity sample.
           The line highlighted in grey is an observation we could drop to keep 
           the observing time under control.
           }
  \label{tab:LSBspec}
  \centering
  \begin{tabular}{|l|l|r|r|}
    \hline
      \textbf{Band [$\boldsymbol{\mu}$m]} & \textbf{Fields} 
        & \textbf{Sensitivity [W/m$^{\boldsymbol{2}}$/sr]} 
        & \textbf{Time [hours]} \\
    \hline
      \textcolor{gray}{Band 1} 
        & \textcolor{gray}{$100\times(2^\prime\times2^\prime)$} 
        & \textcolor{gray}{$6.9\E{-19}$} & \textcolor{gray}{595} \\    
      Band 2 & $100\times(2^\prime\times2^\prime)$ & $1.7\E{-19}$ & 97 \\    
      Band 3 & $100\times(2^\prime\times2^\prime)$ & $1.4\E{-17}$ & 0.6 \\    
      Band 4 & $100\times(2^\prime\times2^\prime)$ & $2.6\E{-17}$ & 0.1 \\    
    \hline
  \end{tabular}
\end{table}


\section{Self-Consistently Probing the Dust Properties and the ISM 
            Structure of Nearby Galaxies}
\label{sec:gasndust}

\subsection{A gas and dust multiphase model}

The degeneracies we face when modeling the dust SEDs of dense regions, and that we have discussed in this paper, originate in our inability to separate the effects due to the variation of the microscopic grain properties and those due to the macroscopic topology of the ISM.
This problem has been addressed, concerning the gas, by state-of-the-art models, where a large number of lines could be used to constrain both the physical conditions of the ISM and its topology in galaxies \cite{cormier12}. 
This has been possible because atomic physics is more precisely known than dust physics. 
The fact that different ions have a wide range of critical densities allows us to use a few well-chosen lines to characterize the main phases of the ISM. 
The knowledge of the average structure of the ISM and of the stellar distribution, provided by the lines, could thus be used as an \textit{a priori} to model the dust properties. 
Yet, obtaining at the same time a well-sampled SED and the intensity of the main IR gas lines is observationally challenging and has been achieved in only a handful of sources, most of the time for a single pointing.

\subsection{Possible observing strategy}

The combined MIR-to-FIR photometric and spectroscopic capability of PRIMA opens the window to obtaining consistent spatially-resolved maps of the total dust and multiphase gas properties in extragalactic regions. 
Such spectra were only obtained in the past by combining ISOSWS and ISOLWS \cite{peeters02}, but with a poor sensitivity, a poor spatial resolution and stitching problems. 
If we have only a few broadband observations, scattered over the whole IR domain, as it is usually the case, we are unable to solve the degeneracy discussed above \cite{galliano18}.
The unique feature brought by PRIMA will be consistent maps of the brightest far-IR lines,
with the well-sampled dust continuum.
\begin{figure}[htbp]
  \includegraphics[width=\textwidth]{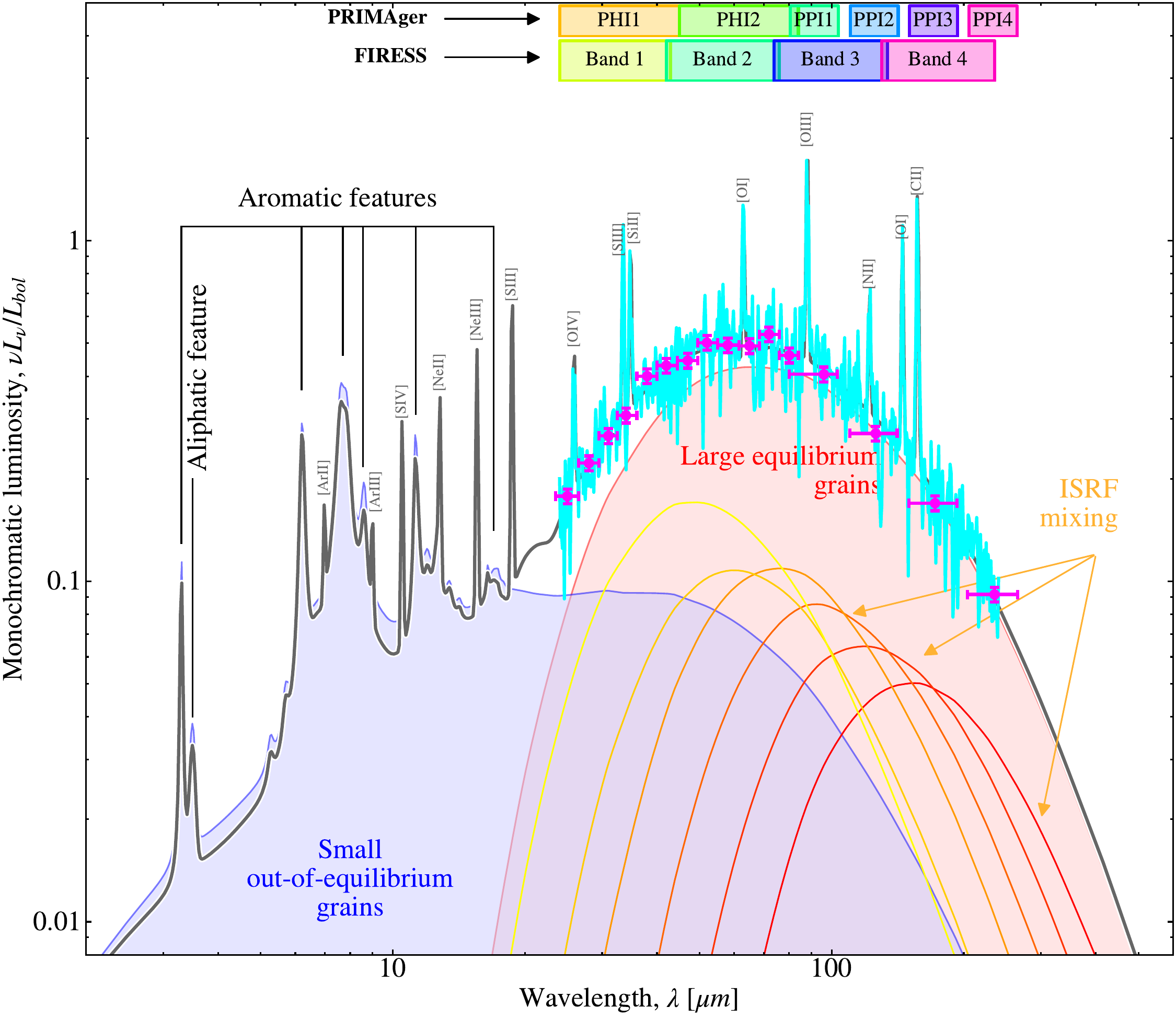}
  \caption{Typical SED of a star-forming region with the brightest gas lines. 
           The magenta error bars correspond to a simulated PRIMAGER broad-band 
           spectrophotometry with $R\simeq4$ for PPI and $R\simeq10$ for PHI ($10\sigma$), 
           and the cyan line is a $R\simeq150$ simulated FIRESS spectrum ($5\sigma$).}
  \label{fig:dustngas}
\end{figure}

We would observe a sample of nearby galaxies, both in narrow-band imaging and low-resolution spectroscopy. 
We do not need to reach the lowest emission of these objects, but we would need to have contiguous maps, in order to understand the spatial variations.
Such combined dust-gas maps could be modeled using \texttt{Cloudy} \cite{ferland17}  and the \texttt{Meudon PDR code} \cite{le-petit06} for the gas, a dust evolution model such as \texttt{THEMIS} \cite{jones17} for the dust and a radiative transfer code such as SOC \cite{juvela19}. 
In addition, recent JWST observations suggest that the PAH emission and the star-formation is enhanced in filamentary structures, which are supposed to have been created by shocks \cite{thilker23}. 
The numerous shock tracers in the FIRESS spectral range could thus be used to understand the heating of the gas in addition to PDRs, using the Paris-Durham shock model \cite{godard19}.
We could observe 30 disk galaxies ($20^\prime\times20^\prime$) and 100 low-metallicity galaxies ($2^\prime\times2^\prime$; 1/10 to 1/3 $Z_\odot$) in order to be able to build an evolutionary sequence of the dust properties.
The outer regions of face-on disk galaxies can also have a low metallicity \cite{chiappini03,gordon08}, and thus complement our sampling of this parameter range.
We would observe this galaxy sample with all PRIMAger bands, and with FIRESS at low spectral resolution.
This could be done in $\simeq860$ hours (\reftabs{tab:dustngas}{tab:dustngaspec}).
\begin{table}[htbp]
  \caption{PRIMAger time estimates for the \refsec{sec:gasndust} science case.}
  \label{tab:dustngas}
  \centering
  \begin{tabular}{|l|*{3}{r|}}
    \hline
      & \textbf{Sensitivity} & \textbf{Disks} & \textbf{Dwarfs} \\
    \hline
      \textbf{PHI1} & 0.85 MJy/sr & 628 hours & 226 hours \\
      \textbf{PHI2} & 4.91 MJy/sr & 5.8 hours & 2.1 hours \\
      \textbf{PPI1} & 14.9 MJy/sr & 0.3 hours & $<0.1$ hours \\
      \textbf{PPI2} & 21.0 MJy/sr & $<0.1$ hours & $<0.1$ hours \\
      \textbf{PPI3} & 18.4 MJy/sr & $<0.1$ hours & $<0.1$ hours \\
      \textbf{PPI4} & 12.3 MJy/sr & $<0.1$ hours & $<0.1$ hours \\
    \hline
  \end{tabular}
\end{table}
\begin{table}[htbp]
  \caption{FIRESS time estimates for the \refsec{sec:gasndust} science case.}
  \label{tab:dustngaspec}
  \centering
  \begin{tabular}{|lr|*{3}{r|}}
    \hline
      & & \textbf{Sensitivity} & \textbf{Disks} & \textbf{Dwarfs} \\
    \hline
      \textbf{Band 1} & \siiilineb & $1.15\E{-16}\;\textnormal{W/m}^2$ 
        & 0.4 hours & 0.2 hours \\
                      & \siIIline & $4.19\E{-17}\;\textnormal{W/m}^2$ 
        & 3.3 hours & 1.2 hours \\
    \hline
      \textbf{Band 2} & \oiline & $5.98\E{-16}\;\textnormal{W/m}^2$ 
        & $<0.1$ hours & $<0.1$ hours \\
    \hline
      \textbf{Band 3} & \oiiiline & $1.69\E{-15}\;\textnormal{W/m}^2$ 
        & $<0.1$ hours & $<0.1$ hours \\
                      & \niiline & $3.82\E{-17}\;\textnormal{W/m}^2$ 
        & $<0.1$ hours & $<0.1$ hours \\
    \hline
      \textbf{Band 4} & \oilineb & $1.02\E{-16}\;\textnormal{W/m}^2$ 
        & $<0.1$ hours & $<0.1$ hours \\
                       & \ciiline & $2.94\E{-15}\;\textnormal{W/m}^2$ 
        & $<0.1$ hours & $<0.1$ hours \\
    \hline
  \end{tabular}
\end{table}

\subsection{Assumptions}

The flux sensitivity has been estimated using the THEMIS model \cite{jones17} (\refapp{app:sens}).
\begin{itemize}
  \item We have assumed $U=3$ and $N(\textnormal{\hi})=10^{21}$~H/cm$^2$ in 
    \reftab{tab:dustngas}.
    This corresponds to the typical extended emission of disk galaxies found in the 
    DustPedia sample \cite{davies17,clark18}. 
    This is however not the most diffuse emission of the galaxy, but this value is 
    sufficient for the present science case, where we are interested in the multiphase 
    nature of the ISM.
  \item The fine-structure lines are all bright. 
    In addition, we do not need to spectrally resolve them. 
    Low-resolution spectroscopy should thus be sufficient to measure their intensity. 
    We have estimated the line intensity in \reftab{tab:dustngaspec}, by assuming they are 
    proportional to the total IR luminosity. 
    The proportionality factor of each line is the average of the sample in Cormier \etal\ 
    (2019)\cite{cormier19}. 
\end{itemize}


\section{Summary and Conclusion}
\label{sec:concl}

To summarize our thoughts, the unprecedented sensitivity of PRIMA will allow us to conduct a few large programs of observations (less than 3000 hours) that should give us key constraints to unlock our understanding of interstellar dust evolution.
\begin{enumerate}
  \item The exceptional sensitivity of PRIMA will allow us to get deep-enough maps of
    external galaxies to detect their most diffuse ISM over the whole MIR-to-FIR window.
    This is essential to build rigorously-constrained extragalactic dust models 
    (\refsec{sec:diffuse}).
  \item PRIMA's unique $\lambda=24-325\emic$ continuous $R=150$ spectroscopic capability 
    will allow us to detect faint solid-state features that are a key to refining 
    our understanding of the chemical composition and structure of interstellar grains
    (\refsec{sec:mineralogy}).
  \item PRIMA's sensitivity will also allow us to measure the FIR SED of quiescent 
    low-metallicity galaxies and thus understand which grain properties are primarily 
    determined by the elemental evolution of the ISM and which one are controlled by the 
    star-formation activity (\refsec{sec:LSB}).
  \item Finally, the continuous spectroscopic capability of PRIMA will allow us to map,
    within the same observations, the MIR-to-FIR SED and numerous gas lines.
    This will allow us to constrain self-consistent multiphase ISM models, and thus 
    solve the degeneracies between the microscopic grain properties and their macroscopic
    topology (\refsec{sec:gasndust}).
\end{enumerate}

These science cases show that PRIMA is a key facility to progress both in our understanding of the physics of the ISM and of galaxy evolution.
Our poor knowledge of the dust properties, a consequence of their complexity, is already the main limitation of the current models of the ISM \cite{rollig07} and galaxy evolution \cite{dubois24}.
Without the observations proposed in this paper, our progress in the major open questions listed in \refsec{sec:intro} would be greatly limited.
The FIR range is crucial to these topics and the possible non-selection of PRIMA by NASA would close the FIR window for several decades.


\appendix    

\section{PRIMA Time Estimation}
\label{app:sens}

  \subsection{Surface brightness estimates}

\begin{figure}[htbp]
  \includegraphics[width=\textwidth]{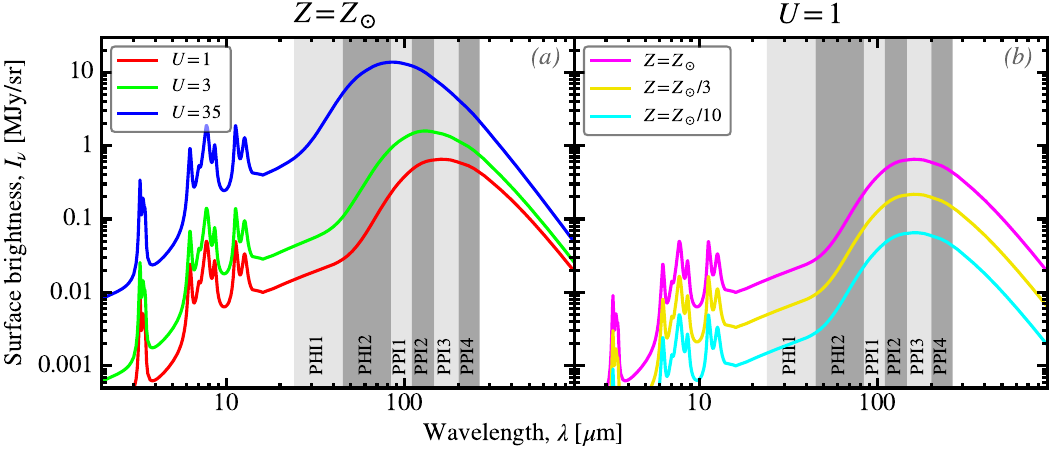}
  \caption{Both panels show the surface brightness of the THEMIS dust model,
           varying the ISRF intensity, $U$ (panel \textit{a}), and varying the 
           metallicity, $Z$ (panel \textit{b}).
           All these curves correspond to the very diffuse ISM, 
           $N(\textnormal{H})=10^{20}\;\textnormal{H/cm}^2$.}
  \label{fig:themis}
\end{figure}

Our surface brightness estimates are based on the THEMIS model \cite{jones17}.
We use it to estimate the surface brightness, $I_\nu$, corresponding to the thermal dust emission, for a given Hydrogen column density, $N_\sms{H}$, and ISRF intensity, $U$.
\begin{description}
  \item[The ISRF] of the Solar neighborhood \cite{mathis83} is simply scaled by the 
    dimensionless parameter $U$ \cite{draine07}, $U=1$ being the value used for the 
    diffuse Galactic ISM.
    The effect of this parameter on the dust surface brightness is non trivial 
    (\reffig{fig:themis}).
  \item[The dust column density] is assumed to scale linearly with metallicity, $Z$: 
    $N(\textnormal{dust})=N(\textnormal{H})\times m_\sms{H}\times Y_\sms{H}\times Z$, where 
    $Y_\sms{H}\equiv M_\sms{dust}/M_\sms{H}=7.4\E{-3}$ 
    and $m_\sms{H}$ is the mass of an H atom \cite{jones17}.
    The dust surface brightness subsequently scales linearly with $N(\textnormal{dust})$.
\end{description}

  \subsection{Observing time}

Our observing strategy is based on the PRIMA instrument characteristics and sensitivities that were available in December 2024 at: \href{https://prima.ipac.caltech.edu/page/instruments}{https://prima.ipac.caltech.edu/page/instruments}.
\begin{description}
  \item[PRIMAger:] the sensitivities, $I_\nu^0$ and $P_\nu^0$, are summarized in 
     \reftab{tab:sens}.
     We thus estimate our observing times, including overheads, for a mapping area $A$ as:
     \begin{eqnarray}
       t_\sms{obs}^\sms{total} & = & \left(\frac{I_\nu^0}{I_\nu}\right)^2 
                                    \times\left(\frac{N_\sigma}{5}\right)^2
                                    \times \frac{A}{1^{\circ2}}
                                    \times 10\;\textnormal{hours} \\
       t_\sms{obs}^\sms{pola} & = & \left(\frac{P_\nu^0}{f_\sms{pol}I_\nu}\right)^2 
                                    \times\left(\frac{N_\sigma}{5}\right)^2
                                    \times \frac{A}{1^{\circ2}}
                                    \times 10\;\textnormal{hours},
     \end{eqnarray}
     where $N_\sigma$ is the number of $\sigma$ we aim to reach
     and $f_\sms{pol}=10\,\%$ is the estimated polarization fraction.
  \item[FIRESS:] the observing time for measuring the integrated intensity, $I$,
    of a line centered at $\lambda$, at level $N_\sigma$, mapped over an area $A$, is:
    \begin{eqnarray}
      t_\sms{obs}^\sms{low-res} & = & 
      \left(\frac{3\E{-19}\;\textnormal{W/m}^2}{I}\right)^2
                                 \times \left(\frac{N_\sigma}{5}\right)^2
                                 \times \frac{A}{100^{\prime2}}
       \nonumber\\
       & & 
       \times \left\{
        \begin{array}{ll}
          800\;\textnormal{hours} 
          & \displaystyle\mbox{for }\;24\emic<\lambda<75\emic \\
          \displaystyle\left(\frac{\lambda}{100\emic}\right)^{-1.68}
          \times 336\;\textnormal{hours}
          & \displaystyle\mbox{for }\;75\emic<\lambda<235\emic \\
        \end{array}
      \right.
    \end{eqnarray}
\end{description}
\begin{table}[htbp]
  \caption{PRIMAger extended source sensitivities for a $5\sigma$ background-subtracted 
           1 square degree map observed during 10 hours.}
  \label{tab:sens}
  \centering
  \begin{tabular}{|l|*{6}{r|}}
    \hline
      & \textbf{PHI1} & \textbf{PHI2} & \textbf{PPI1} & \textbf{PPI2} & \textbf{PPI3} 
      & \textbf{PPI4} \\
    \hline
      \textbf{Total power}, $I_\nu^0$ [MJy/sr] & 4.5 & 2.5 & 1.58 & 1.20 & 0.88 & 0.65 \\
      \textbf{Polarized intensity}, $P_\nu^0$ [MJy/sr] & \ldots & \ldots 
         & 2.23 & 1.70 & 1.25 & 0.91 \\
    \hline
  \end{tabular}
\end{table}
\begin{table}[htbp]
  \caption{FIRESS low-resolution parameters.}
  \label{tab:}
  \centering
  \begin{tabular}{|l|*{4}{r|}}
    \hline
      & \textbf{Band 1} & \textbf{Band 2} & \textbf{Band 3} & \textbf{Band 4} \\
    \hline
      \textbf{Wavelength range} & 24--43 \tmic & 42--76 \tmic & 74--134 \tmic & 130--235 \\
      \textbf{Wavelength sampling} & 0.23 \tmic & 0.41 \tmic & 0.73 \tmic & 1.29 \tmic \\
      \textbf{Slit width}, $W(\lambda)$ & $7.6^{\prime\prime}$ & $7.6^{\prime\prime}$ 
        & $12.7^{\prime\prime}$ & $22.9^{\prime\prime}$ \\
    \hline
  \end{tabular}
\end{table}

\subsection*{Disclosures}
The authors do not have any conflict of interest to disclose.

\subsection*{Code, Data, and Materials Availability} 
All the simulations in this paper were performed using the \texttt{THEMIS} dust model that can be freely downloaded from \href{https://www.ias.u-psud.fr/DUSTEM/}{https://www.ias.u-psud.fr/DUSTEM/}.

\subsection*{Acknowledgments}
We thank our two anonymous referees for their insightful comments.
This work is supported by the French National Research Agency under the contract WIDENING ANR-23-ESDIR-0004, as well as by the Programme National ``Physique et Chimie du Milieu Interstellaire'' (PCMI) of CNRS/INSU, with INC and INP, and the ``Programme National Cosmologie et Galaxies'' (PNCG) of CNRS/INSU, with INP and IN2P3, both programs being co-funded by CEA and CNES.


\bibliography{references}   
\bibliographystyle{spiejour}   


\vspace{2ex}\noindent\textbf{Frédéric Galliano} is a staff researcher at CNRS/AIM associated with the Department of Astrophysics of CEA Paris-Saclay. 
He received his MS and PhD degrees in astrophysics from the University of Paris XI in 2000 and 2004, respectively.  
He is the author of about 200 journal papers. 
He develops the hierarchical Bayesian dust SED code HerBIE.
His current research interests include interstellar dust grains, and infrared and millimeter observations.

\vspace{2ex}\noindent\textbf{Simone Bianchi} is associate researcher at the INAF/Arcetri Astrophysical Observatory in Florence. 
He received his Laurea degree in physics from the University of Florence in 1995 and PhD degree in astrophysics from the University of Wales, College of Cardiff, in 1999. 
He is the author of about 200 journal papers. His current research interests include 
interstellar dust grains and millimeter/radio observations.

\vspace{2ex}\noindent\textbf{Mika Juvela} is a university lecturer at the department of physics, University of Helsinki. 
He got his PhD degree in astronomy from the University of Helsinki in 1997. 
He is the author of more than 300 papers, and his current research interests include infrared and radio astronomy, Galactic star formation, and radiative transfer.

\vspace{2ex}\noindent\textbf{Hidehiro Kaneda} is a professor at Nagoya University. 
He received his MS and PhD degrees in X-ray astrophysics from the University of Tokyo in 1994 and 1997, respectively. 
He was the deputy PI of the former far-IR satellite project SPICA. 
His current research interests include interstellar dust grains and cryogenic optics for space IR missions.

\vspace{2ex}\noindent\textbf{Vianney Lebouteiller} is a staff researcher at CNRS/AIM associated with the Department of Astrophysics of CEA Paris-Saclay. 
He received his MS and PhD degrees in astrophysics from the University of Paris VI in 2001 and 2005, respectively. 
He is PI of the CASSISjuice \spitz\ spectral atlas. 
His current research interests include extremely metal-poor galaxies and physical models of the interstellar medium.

\vspace{2ex}\noindent\textbf{Takashi Onaka} is a professor emeritus at the University of Tokyo. He received his BS, MS, and PhD degrees in astronomy from the University of Tokyo in 1975, 1977, and 1980, respectively.
He is the author of more than 300 journal papers. 
He was the PI of Infrared Camera (IRC) on board the Japanese infrared satellite AKARI. 
His current research interests include interstellar dust grains and infrared observations.

\vspace{2ex}\noindent\textbf{Lara Pantoni} is a Postdoc at the University of Ghent. 
She received her BS and MS degrees in astronomy from the University of Bologna in 2014 and 2017, respectively. 
She received her PhD degree in astrophysics from SISSA in 2021. She is the author of about 20 journal papers.  
Her current research interests include interstellar dust grains in local galaxies and AGNs, dusty galaxies at the Cosmic Noon, infrared and millimeter
observations.

\vspace{2ex}\noindent\textbf{Tsutomu T.\ Takeuchi} is an Associate Professor at Nagoya University, Japan. 
He received his BS, MS, and Sc.D.\ (equivalent to a PhD) in astrophysics from Kyoto University in 1994, 1997, and 2000, respectively. 
He has authored over 400 journal papers. He contributed to the Japanese infrared satellite AKARI. 
His research focuses on the formation and evolution of galaxies with dust physics, as well as interdisciplinary studies bridging astrophysics, cosmology, and data science.

\vspace{1ex}\noindent Biographies and photographs of the other authors are not available.

\listoffigures
\listoftables

\end{spacing}
\end{document}